\documentclass[preprint,12pt]{elsarticle}

\usepackage{amssymb}
\usepackage{bm}
\usepackage{hyperref}
\usepackage{amsmath}

\usepackage{lineno}

\journal{Computer Methods and Programs in Biomedicine}

\begin{document}

\begin{frontmatter}

\title{Data-driven generation of 4D velocity profiles in the aneurysmal ascending aorta}

\author[poli]{Simone Saitta}
\author[poli,imperial]{Ludovica Maga}
\author[imperial]{Chloe Armour}
\author[poli]{Emiliano Votta}
\author[mrc]{Declan P. O'Regan}
\author[s&c]{M. Yousuf Salmasi}
\author[s&c]{Thanos Athanasiou}
\author[cornell]{Jonathan W. Weinsaft}
\author[imperial]{Xiao Yun Xu}
\author[imperial,delft]{Selene Pirola}
\ead{s.pirola@tudelft.nl}
\author[poli]{Alberto Redaelli}

\affiliation[poli]{organization={Department of Information, Electronics and Bioengineering, Politecnico di Milano},
            city={Milan},
            country={Italy}}

\affiliation[imperial]{organization={Department of Chemical Engineering, Imperial College London},
            city={London},
            country={UK}}
            
\affiliation[mrc]{organization={MRC London Institute of Medical Sciences, Imperial College London},
            city={London},
            country={United Kingdom}}
            
\affiliation[s&c]{organization={Department of Surgery and Cancer, Imperial College London},
            city={London},
            country={United Kingdom}}

\affiliation[cornell]{organization={Department of Medicine (Cardiology), Weill Cornell College},
            city={New York, NY},
            country={USA}}
        
\affiliation[delft]{organization={Department of BioMechanical Engineering,
3mE Faculty, Delft University of Technology},
            city={Delft},
            country={Netherlands}}


\begin{abstract}
\textit{Background and Objective}:
Numerical simulations of blood flow are a valuable tool to investigate the pathophysiology of ascending thoratic aortic aneurysms (ATAA). To accurately reproduce \textit{in vivo} hemodynamics, computational fluid dynamics (CFD) models must employ realistic inflow boundary conditions (BCs). However, the limited availability of \textit{in vivo} velocity measurements, still makes researchers resort to idealized BCs. The aim of this study was to generate and thoroughly characterize a large dataset of synthetic 4D aortic velocity profiles with features similar to clinical cohorts
of patients with ATAA.\\
\textit{Methods}:
Time-resolved 3D phase contrast magnetic resonance (4D flow MRI) scans of 30 subjects with ATAA were processed through in-house code to extract anatomically consistent cross-sectional planes along the ascending aorta, ensuring spatial alignment among all planes and interpolating all velocity fields to a reference configuration. Velocity profiles of the clinical cohort were extensively characterized by computing flow morphology descriptors of both spatial and temporal features. By exploiting principal component analysis (PCA), a statistical shape model (SSM) of 4D aortic velocity profiles was built and a dataset of 437 synthetic cases with realistic properties was generated.\\
\textit{Results}:
Comparison between clinical and synthetic datasets showed that the synthetic data presented similar characteristics as the clinical population in terms of key morphological parameters. The average velocity profile qualitatively resembled a parabolic-shaped profile, but was quantitatively characterized by more complex flow patterns which an idealized profile would not replicate. 
Statistically significant correlations were found between PCA principal modes of variation and flow descriptors. \\
\textit{Conclusions}:
We built a data-driven generative model of 4D aortic velocity profiles, suitable to be used in computational studies of blood flow. The proposed software system also allows to map any of the generated velocity profiles to the inlet plane of any virtual subject given its coordinate set.

\end{abstract}


\end{frontmatter}


\section{Introduction}
\label{sec:introduction}
Thoracic aortic aneurysm (TAA) is a life-threatening condition involving an abnormal dilatation of the aortic wall 
\citep{elefteriades2002natural}.
An accurate assessment of blood flow plays an essential role in clinical diagnosis, risk stratification and treatment planning of TAA \citep{catalano2021atlas, yeung2006aortoiliac, chien1998effects}.
Computational fluid dynamics (CFD) is a well established tool to quantify hemodynamics \citep{pirola20194,mendez2018comparison} 
through \textit{in silico} trials \citep{li2005blood, peirlinck2021precision}. 
To achieve a high level of fidelity, CFD models need to account for patient-specific boundary conditions (BCs). When choosing inflow BCs, prescribing patient-specific data in the form of 3-directional velocity profiles allows to obtain significantly more accurate results compared to using idealized profiles, as amply shown by several recent studies that make use of velocity information extracted from phase-contrast magnetic resonance imaging (PC-MRI) \citep{morbiducci2013inflow, pirola2018computational, youssefi2018impact, armour2021influence}. 
Nonetheless, the limited availability of \textit{in vivo} velocity measurements, still makes researchers resort to idealized BCs. Moreover, such lack of clinical data represents an obstacle for setting up population-based \textit{in silico} trials and for building datasets for training machine learning (ML) models.
Generative models can be used to overcome this limitation by creating larger data-driven synthetic datasets \citep{romero2021clinically}. In particular, statistical shape models (SSMs) have been adopted in the cardiovascular field \citep{young2009computational, casciaro2014identifying}.
SSMs are data-driven approaches for assessing shape variability and creating large virtual cohorts from clinical ones. An SSM is typically based on principal component analysis (PCA) and describes the shape probability distribution of the input data by a mean shape and modes of shape variations
\citep{jollife2016principal}. 
Several studies have effectively applied SSMs to study TAA geometry \citep{liang2017machine, thamsen2021synthetic, cosentino2020statistical}. 
Nonetheless, aortic hemodynamics, which have been shown to play a key role in pathophysiology of this disease \citep{mendez2018comparison, nannini2021aortic, salmasi2021high}, have not received the same attention. An exception is the work of Catalano \textit{et al.}, who exploited SSMs to build an atlas of aortic hemodynamics in subjects with by bicuspid (BAV) and tricuspid (TAV) aortic valve \citep{catalano2021atlas}. However, the authors imposed an idealized parabolic velocity profile as inlet BC for their CFD models. Despite revealing important insights on BAV vs. TAV biomarkers, the study is hampered by the use of such simplified inlet BCs, which 
significantly affect the computed aortic blood flow, especially in regions that are close to the inlet, namely the ascending aorta \citep{youssefi2018impact, pirola2017choice, armour2021influence}.\\
Motivated by the need for boosting the impact and the fidelity of numerical studies involving ascending TAA (ATAA) hemodynamics, the present work leverages SSMs to pursue three specific aims.
We provide: i) a quantitative and detailed characterization of a representative 4D ATAA inlet velocity profile as a valid alternative to idealized inlet BCs for numerical simulations; ii) a synthetic virtual cohort of 4D ATAA inlet velocity profiles with features that are consistent with those of real ATAA inlet profiles and potentially large enough to allow for ML approaches to be used; iii) insights into both spatial and temporal hemodynamic features of ATAA velocity fields in the ascending aorta. 


\section{Methods}
\label{sec:methods}


\subsection{Image data}
Thoracic 4D flow MRI scans of 30 subjects with ATAA acquired between 2017 and 2019 were retrospectively retrieved. Our dataset included fully deintentified images provided by Weill Cornell Medicine, (NY, USA) and Hammersmith Hospital (London, United Kingdom). 
None of the subjects in our cohort were BAV-affected. 
Respiratory compensated 4D flow acquisitions were performed 
with the following settings: spatial resolution (voxel size) 1.4 -- 2.0 mm (range), field of view = 360 mm, flip angle = 15°, VENC = 150 -- 200 cm/s (range), 
time resolution 20 -- 28 frames/cardiac cycle (range).
Data usage was approved by the Weill Cornell Medicine Institutional Review Board (New York, NY, USA) and by the Health Research Authority (HRA) (17/NI/0160) in the UK and was sponsored by the Imperial College London Joint Research and Compliance Office, as defined under the sponsorship requirements of the Research Governance Framework (2005). The participants provided their written informed consent to participate in this study.


\subsection{Data preprocessing}
\label{subsec:preproc}
\begin{figure}[ht]
\includegraphics[width=\textwidth]{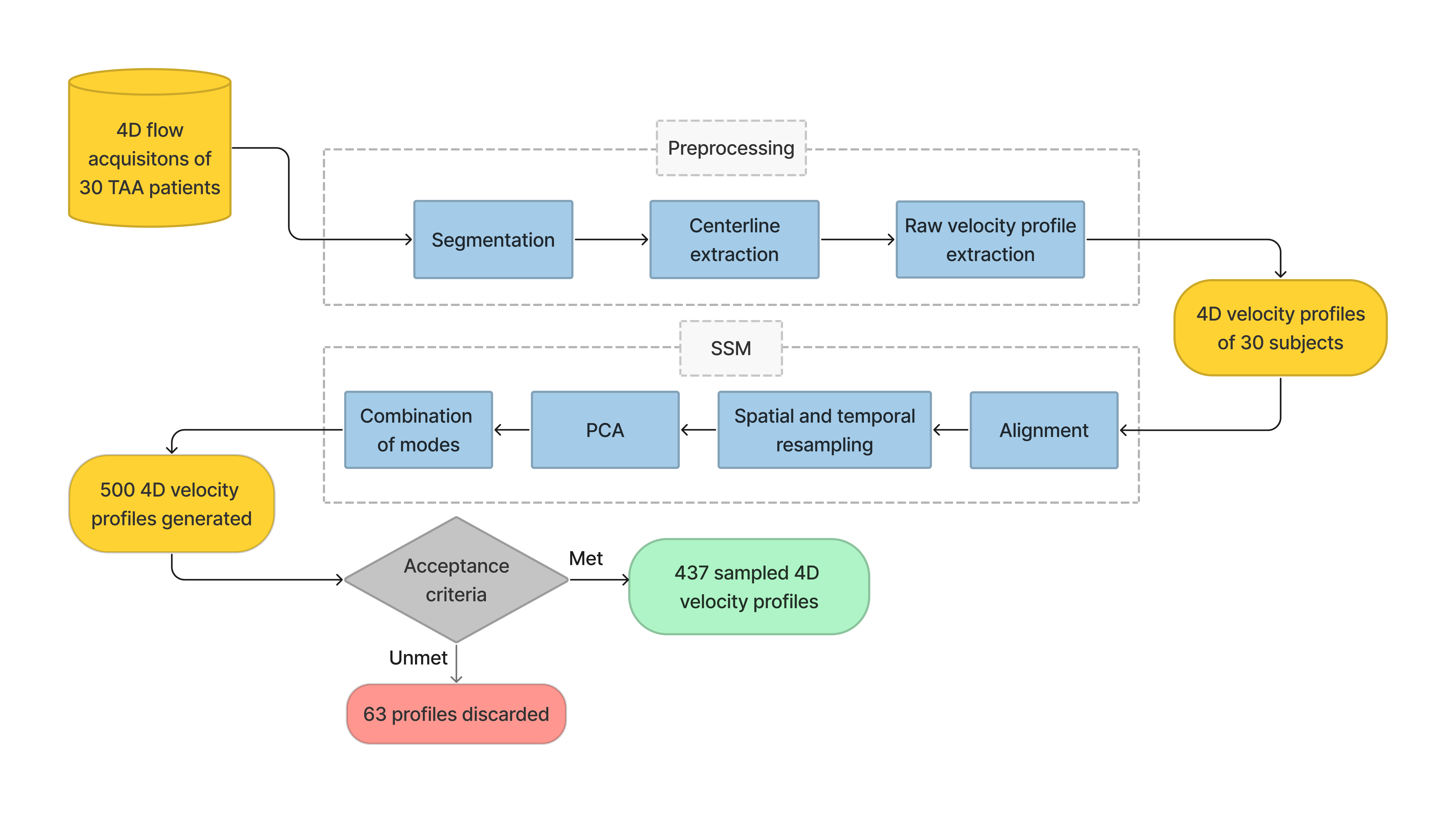}
\caption{Schematic representation of the adopted workflow. All 4D flow acquisitions go through a preprocessing pipeline for the extraction of velocity profiles. The SSM process consists in a common alignment and spatiotemporal resampling of the profiles and then a combination of PCA modes to generate new ones. Only the profiles that meet specific acceptance criteria are included in the final dataset.} 
\label{workflow_fig}
\end{figure}

4D flow MRI data were preprocessed using in-house Python code and following the workflow presented in figure~\ref{workflow_fig}: for each patient, a 3D binary mask of the aorta was extracted from PC-MR angiography (PC-MRA) images using semi-automatic tools available in the open source software ITK-SNAP \citep{yushkevich2017itk}. 
To guarantee consistency of inlet plane location among all ATAA subjects, inlet planes were defined with respect to a commonly used anatomical landmark represented by the bifurcation of the pulmonary artery (PA) \citep{saitta2022deep}. 
A triangulated mesh of the selected plane within the aortic lumen was generated; 4D flow velocity data were then probed at inlet plane nodal locations. For the generic subject indexed by $j$, we defined the inlet plane nodal coordinates as $\tilde{\Xi}^{(j)} = [\tilde{\bm{\xi}}_1^{(j)},..., \tilde{\bm{\xi}}_\tau^{(j)},..., \tilde{\bm{\xi}}_{\mathcal{T}^{(j)}}^{(j)}]^\intercal$, and the corresponding measured velocity vector field as $\tilde{\mathcal{V}}^{(j)} = [\tilde{\bm{v}}_1^{(j)},..., \tilde{\bm{v}}_\tau^{(j)},..., \tilde{\bm{v}}_{\mathcal{T}^{(j)}}^{(j)}]^\intercal$, with 
$\tilde{\bm{\xi}}_\tau^{(j)},  \tilde{\bm{v}}_\tau^{(j)} \in \mathbb{R}^{\mathcal{N}^{(j)} \times 3}$ and where
$\mathcal{T}^{(j)}$ and $\mathcal{N}^{(j)}$ are the number of frames in the cardiac cycle of subject $j$ and the number of probed nodal locations on the inlet plane, respectively; therefore, in general, the dimensions of $\tilde{\Xi}^{(j)}$ and $\tilde{\mathcal{V}}^{(j)}$ vary among subjects.

\subsection{Statistical shape modeling}\label{subsec:ssm}

\paragraph{Alignment}
Consistent spatial orientation among the extracted inlet velocity profiles was ensured through two steps: first, each inlet plane was centered at the origin by applying the translation $\mathbf{T}^{(j)}$ to transform the nodal coordinates $\tilde{\bm{\xi}}_\tau^{(j)}$ into $\bm{\tilde{x}}_\tau^{(j)} = \tilde{\bm{\xi}}_\tau^{(j)} + \mathbf{T}^{(j)}$. Second, two consecutive rigid rotations were applied to the translated coordinates $\tilde{\bm{x}}_\tau^{(j)}$ and to the corresponding velocities $\tilde{\bm{v}}_\tau^{(j)}$. The first rotation ($\mathbf{R}_1 \in \mathbb{R}^{3 \times 3}$) transformed $\tilde{\bm{x}}_\tau^{(j)}$ and the corresponding velocity profile $\tilde{\bm{v}}_\tau^{(j)}$ to $\hat{\bm{x}}_\tau^{(j)} = \mathbf{R}_1^{(j)} \tilde{\bm{x}}_\tau^{(j)}$ and $\hat{\bm{v}}_\tau^{(j)} = \mathbf{R}_1^{(j)} \tilde{\bm{v}}_\tau^{(j)}$, respectively, so to make the inlet plane containing $\hat{\bm{x}}_\tau^{(j)}$ normal to the $z$-axis. The second rigid rotation ($\mathbf{R}_2 \in \mathbb{R}^{3 \times 3}$) transformed $\hat{\bm{x}}_\tau^{(j)}$ and $\hat{\bm{v}}_\tau^{(j)}$ to $\bm{x}_\tau^{(j)} = \mathbf{R}_2^{(j)} \hat{\bm{x}}_\tau^{(j)}$ and $\bm{v}_\tau^{(j)} = \mathbf{R}_2^{(j)} \hat{\bm{v}}_\tau^{(j)}$, and it ensured that the $x$-axis was aligned with the right-to-left direction of the subject.

\begin{figure}
    \centering
    \includegraphics[width=0.9\textwidth]{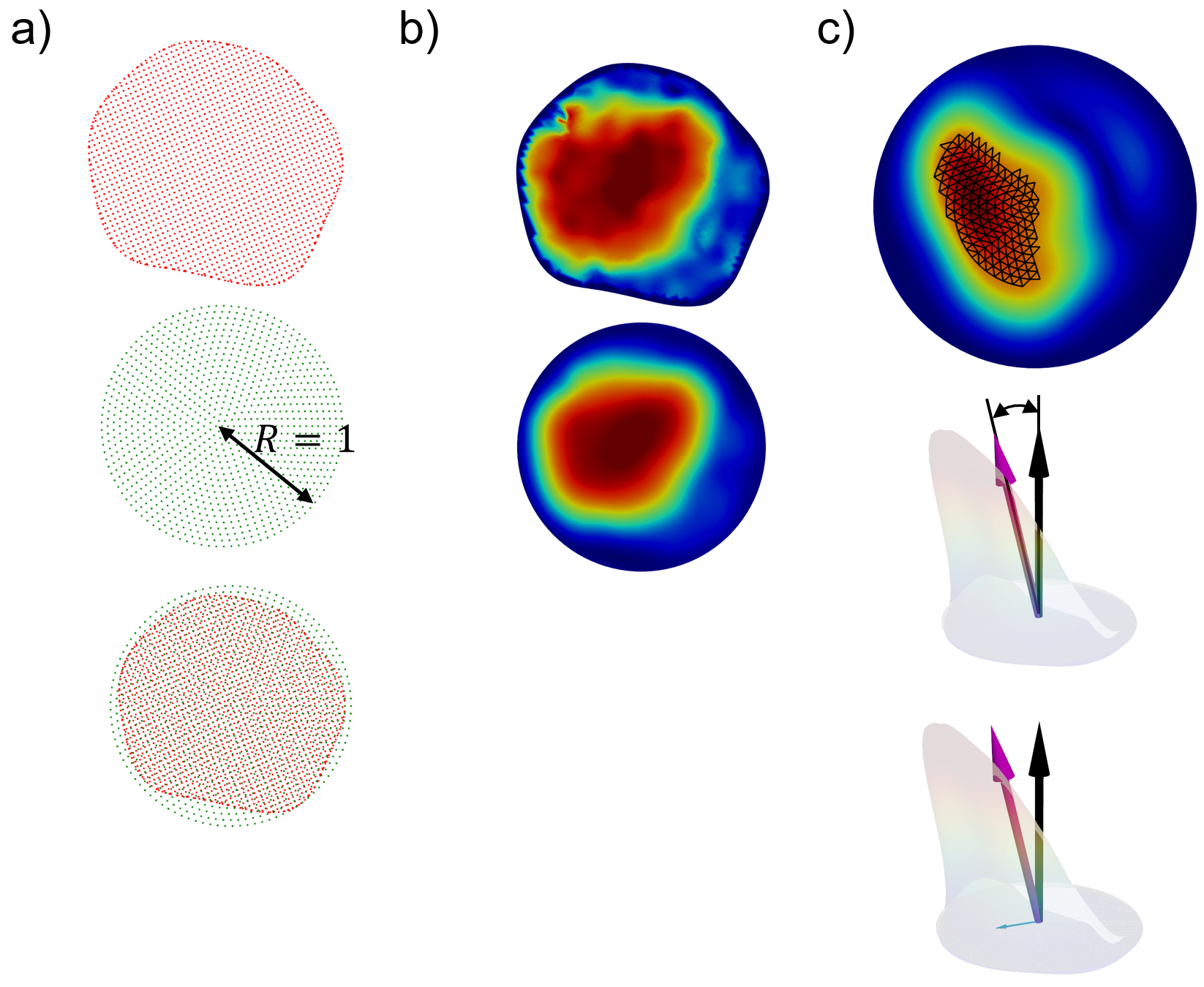}
    \caption{a) Representation of an inlet plane coordinate set from a representative subject in the clinical cohort (red), as obtained upon rigid roto-translation, and of the corresponding point set in the fixed disk domain (green). b) A velocity profile from the clinical cohort (top), resampled to the fixed reference disk using RBF (bottom). c) Exemplification of the computed flow descriptors. FDI (top): $area^{top15\%}$ is highlighted in black; FJA (middle): angle between plane normal (black) and mean velocity direction (magenta); SFD (bottom): ratio between the normal component (black) and in-plane component (cyan) of the mean velocity vector (magenta).}
    \label{fig:flow_descr}
\end{figure}

\paragraph{Resampling}
After alignment, each velocity profile $\bm{v}_\tau^{(j)}$ was mapped onto a reference disk with unit radius using linear radial basis functions, effectively enabling the resampling of each velocity profile at $N=1071$ fixed spatial locations uniformly distributed over the reference disk (figure \ref{fig:flow_descr}a and b).
Each velocity profile time sequence was temporally interpolated to a reference temporal interval $t \in [0, 1]$ discretized in $T=20$ frames, using cubic polynomials. 
Finally, for the generic subject $j$, the spatiotemporally aligned and resampled velocity profiles are defined as: 
$\mathbf{V}^{(j)} = [\mathbf{v}_1^{(j)},..., \mathbf{v}_t^{(j)},..., \mathbf{v}_{T}^{(j)}]^\intercal$, with 
$\mathbf{v}_t^{(j)} \in \mathbb{R}^{N \times 3}$.

\paragraph{Principal component analysis}
The 30 aligned 4D velocity profiles were rearranged into column vectors and assembled into a matrix $\mathbf{V} = [ \mathbf{V}^{(1)},\allowbreak ...,\allowbreak \mathbf{V}^{(j)}, ..., \mathbf{V}^{(30)} ]$, with $\mathbf{V} \in \mathbb{R}^{P \times J}$ where $J$ is the number of subjects (30) and $P=3 \times N \times T$. 
Matrix $\mathbf{V}$ was used as input for a PCA. 
Standard PCA begins by computing the mean velocity profile defined as:
\begin{equation}\label{eq_mean}
    \bar{\mathbf{V}} = \frac{1}{J} \sum_{j=1}^{J} \mathbf{V}^{(j)},  
\end{equation}
and by assembling the covariance matrix $\mathbf{C}$, given by:
\begin{equation}\label{eq_cov}
    \mathbf{C} = \frac{1}{J} \sum_{j=1}^{J} (\mathbf{V}^{(j)} - \bar{\mathbf{V}}) (\mathbf{V}^{(j)} - \bar{\mathbf{V}}^\intercal).  
\end{equation}
The eigenvalues and eigenvectors of $\mathbf{C}$ were sorted in descending order to obtain a sequence of eigenvectors that progressively maximize the explained variance. The first 18 eigenvectors were considered since their cumulative variance was $\geq 90\%$. Each mode represents a shape direction of variation from the mean velocity profile that is representative of data variability.

\paragraph{Shape sampling}
Starting from the mean shape $\bar{\mathbf{V}}$, the SSM can be built, and a synthetic 4D profile $\mathbf{U}$ can be generated by adding to $\bar{\mathbf{V}}$ a shape variation, i.e., a linear combination of the selected modes \citep{liang2017machine} as:

\begin{equation}\label{eq_ssm}
    \mathbf{U} = \bar{\mathbf{V}} + \sum_{m=1}^{M} b^{(m)} \sqrt{\lambda^{(m)}} \bm{a}^{(m)}
\end{equation}

where $M$ denotes the number of selected modes, $\bm{a}^{(m)}$ is the  eigenvector of $\mathbf{C}$ associated to the $m$-th selected mode, and $\lambda^{(m)}$ is the corresponding eigenvalue. New profile shapes can be sampled from the SSM by using a set of coefficients, or weights, $\bm{b} = [b^{(1)}, ..., b^{(m)}, ..., b^{(M)}]$. In particular, to study the shape variations captured by a specific mode, one can sample shapes by considering only the selected mode and varying the coefficient $b^{(m)}$.\\ 
A uniform sampling technique for the first 18 modes of variation of the PCA was used to generate the virtual dataset of inlet velocity profiles: 
\begin{equation}\label{eq_sampling}
    b^{(m)} \sim \mathcal{U}(-1.5, 1.5).
\end{equation}


\subsection{Flow morphology descriptors}\label{subsec:descriptors}
To quantitatively characterize inlet velocity profiles belonging to the clinical cohort and to the synthetic one, several descriptors of flow morphology were computed, both at systolic peak and as time-averaged quantities: positive peak velocity (PPV), flow dispersion index (FDI), flow jet angle (FJA), secondary flow degree (SFD), 
and retrograde flow index (RFI). Representations of FDI, FJA and SFD is included in Figure \ref{fig:flow_descr}c. For a generic profile $\bm{v}_t^{(j)}$ consisting of nodal velocity vectors $\bm{v}_{t,n}^{(j)}$, with $n=1,...,N$, flow descriptors were computed as follows.

\begin{description}

\item[Flow dispersion (FDI)] was calculated to determine whether flow displayed a broad or peaked in-plane distribution. FDI was computed as the ratio between the area of the region characterized by the top 15\% of peak velocity magnitudes ($area^{top15\%}$) and the inlet area \citep{youssefi2018impact}:
\begin{equation}\label{eq_fdi}
    FDI = \frac{area^{top15\%}}{area} \times 100 \%, 
\end{equation}
Accordingly, the higher the FDI value the more homogeneous the velocity profile; the lower the FDI value the sharper the velocity profile.

\item[Flow jet angle (FJA)] represents the angle formed by the mean velocity direction
(jet direction $\bm{v}_{mean}$) and the unit vector orthogonal to the inlet surface $\bm{n}$:
\begin{equation}\label{eq_fja}
    FJA = arccos(\bm{v}_{mean} \cdot \bm{n}), 
\end{equation}
FJA quantifies the skewness of the inlet flow towards aortic walls. An FJA value of 0\textdegree represents a mean jet direction perpendicular to the inlet plane.

\item[Secondary flow degree (SFD)] 
is computed as the ratio between the mean in-plane (radial) velocity magnitude $\bm{v}_\parallel$ and the mean axial velocity magnitude (through-plane velocity) $\bm{v}_\bot$ as:
\begin{equation}\label{eq_sfd}
    SFD = \frac{||\bm{v}_\parallel||}{||\bm{v}_\bot||}, 
\end{equation}


\item[Retrograde flow index (RFI)] was calculated as the fraction of negative area under the curve of the flow rate time-course over the whole area under the curve \citep{saitta2021qualitative}:

\begin{equation}
    RFI=\frac{\left|\int_{0}^{T} Q_{r} d t\right|}{\left|\int_{0}^{T} Q_{a} d t\right|+\left|\int_{0}^{T} Q_{r} d t\right|} \times 100 \%,
\end{equation}
where $Q_r$ and $Q_a$ are the total retrograde and antegrade flow rate, respectively. A higher RFI value implies an increasing flow direction inversion during the cardiac cycle.

\end{description}


\subsection{Acceptance criteria}\label{subsec:acceptance}
To avoid generation of unrealistic velocity profiles and restrict the synthetic population only to plausible cases
acceptance criteria were introduced 
Such criteria were based on the flow features of the clinical profiles. Specifically, intervals of acceptance $I_d$ were defined based on the statistical distributions of the flow descriptors defined in section~\ref{subsec:descriptors} and computed as:
\begin{equation}
    I_{d}=\left[\mu_{d}-2 \sqrt{\lambda_{d}}\right] \cup\left[\mu_{d}+2 \sqrt{\lambda_{d}}\right]
\end{equation}
where $\mu_d$ denotes the mean value of the considered flow descriptor $d$:
\begin{equation}
    \mu_d = \frac{1}{J} \sum_j \frac{1}{T} \sum_t d_t^{(j)},
\end{equation}
and $\sqrt{\lambda_d}$ its standard deviation.
Flow descriptors were extracted from the synthetic velocity profiles; those synthetic profiles characterized by at least one parameter falling outside the acceptance intervals were automatically rejected. 

\subsection{Statistical analysis} \label{subsec:statistics}
To assess statistical differences between the clinical and synthetic sets, comparisons were made using unpaired t-tests for normally distributed variables and Mann–Whitney U tests for non-normally distributed data. Data normality was determined using the Shapiro–Wilk test.\\
Pearson's correlation coefficients were calculated to assess correlations between shape modes and flow descriptors. \textit{p}-values \textless{0.05} were considered statistically significant.


\section{Results}
\label{sec:results}

The devised SSM was exploited to analyze the modes of variation and assess their correlation with flow morphological features. By containing the majority of statistical information, the first four modes were responsible for approximately $45\%$ of the total dataset variability (figure~\ref{profiles_fig}a). 

\subsection{Statistical shape model analysis}
\begin{figure}[ht]
\centering
\includegraphics[width=\textwidth]{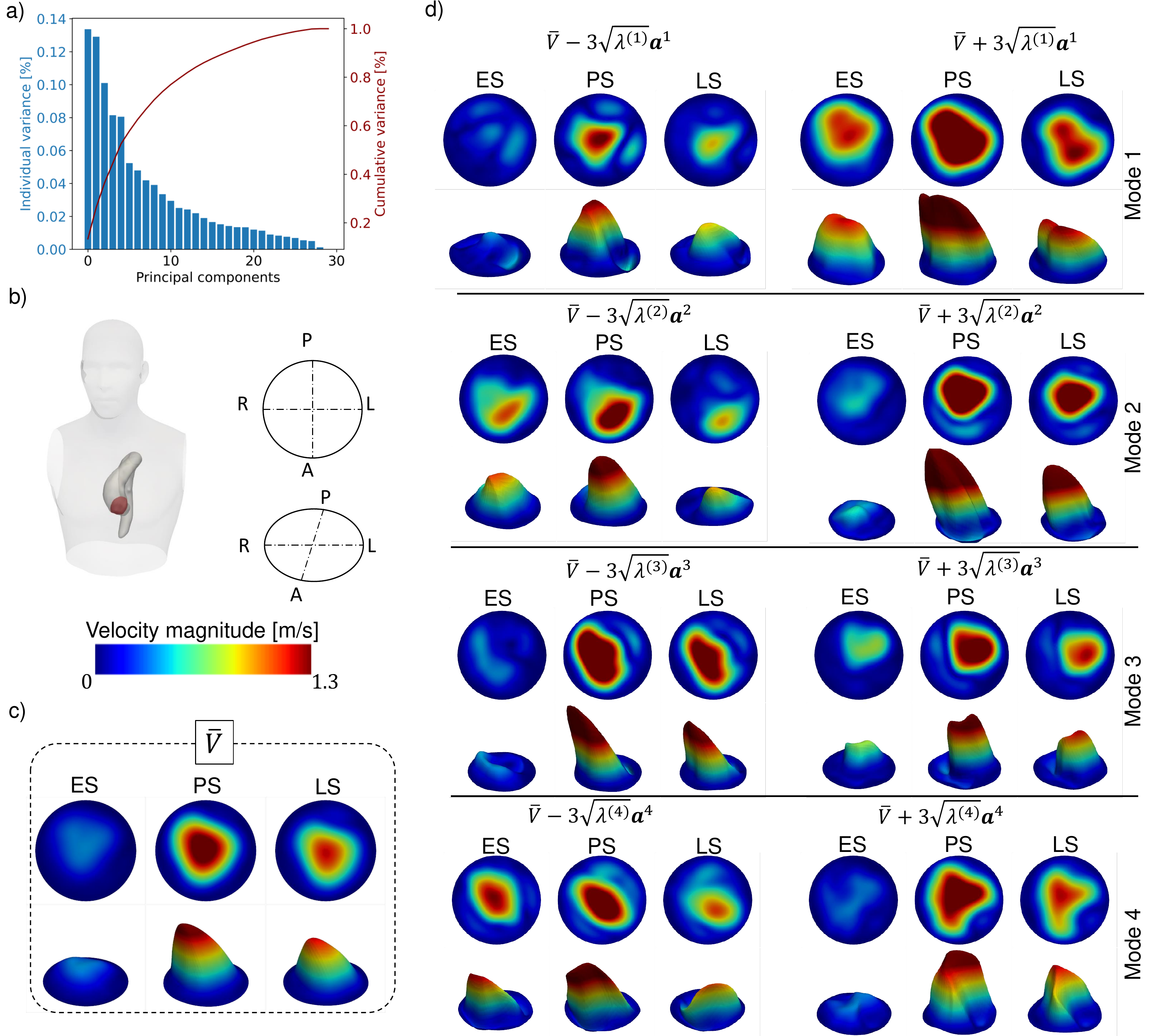}
\caption{a. Individual and cumulative variance associated with each principal component. b. Orientation of the displayed planes and profiles in with respect to the subject in 2D and 3D. c. Mean velocity profile ($\bar{\mathbf{V}}$) colored by velocity magnitude in 2D (top row) and 3D (bottom row) at early systole (ES), peak systole (PS) and late systole (LS). d. 2D and 3D visualizations of velocity profiles deformed towards minimum and maximum for the first 4 modes and colored by velocity magnitude.} 
\label{profiles_fig}
\end{figure}

\paragraph{Characterization of the clinical cohort} For the original cohort, the mean 4D profile $\bar{\mathbf{V}}$ was obtained and characterized in terms of flow descriptors. Time-averaged descriptors for the mean profile of the clinical cohort were: $PPV=0.42\: m/s$, $FDI=13.34\%$, $FJA=0.42^{\circ}$ and $SFD=7.57$. Values at peak systole (PS) were: $PPV=1.47\: m/s$, $FDI=12.8\%$, $FJA=13^{\circ}$, $SFD=0.23$ 
;while $RFI=2\%$. 
$\bar{\mathbf{V}}$ (orientated as in figure \ref{profiles_fig}b) can be visualized at three key time points through the cardiac cycle: early systole (ES), PS and late systole (LS) (figure \ref{profiles_fig}c).

\paragraph{Generation of the synthetic inlet velocity profiles} The sampling of the SSM through the process described in section \ref{subsec:ssm} led to generate 500 synthetic 4D profiles. Out of these, the acceptance criteria led to accepting 437 velocity profiles that constituted the final synthetic cohort.

\begin{figure}[ht]
\centering
\includegraphics[width=\textwidth]{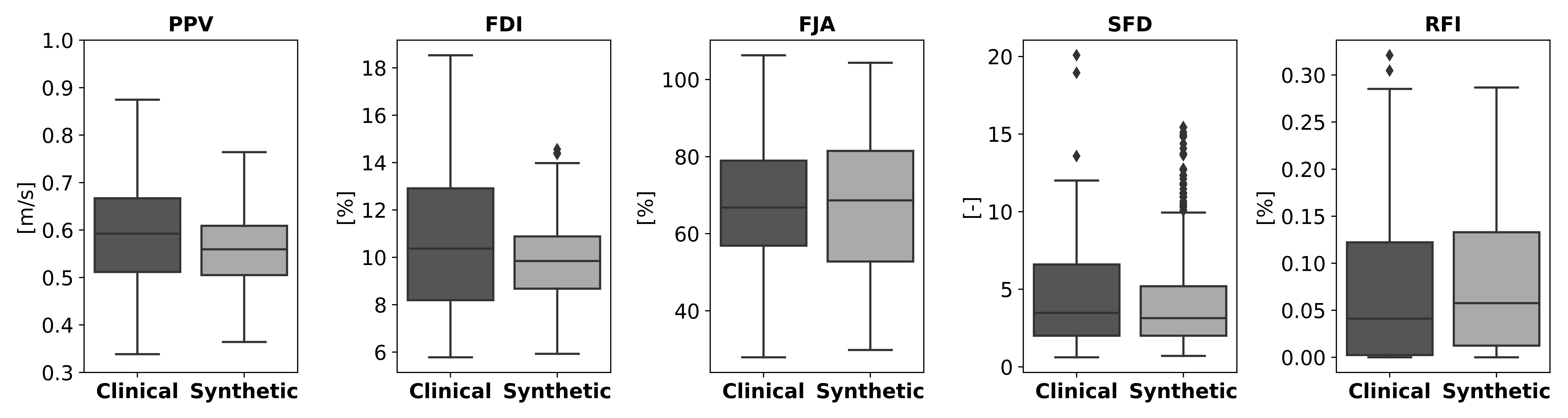}
\caption{Box plots showing distributions of time-averaged flow descriptors for real and synthetic cohorts.
Similar medians and ranges were observed. No significant differences ($p \geq 0.05$) were found except for PPV ($p=0.040$). Whiskers indicate 1.5 interquartile ranges; diamonds highlight outliers.} 
\label{boxplot_fig}
\end{figure}

\paragraph{Comparison of synthetic vs. clinical velocity profiles} Synthetic and clinical inlet velocity profiles were compared based on time-averaged flow descriptors shown in table \ref{tab:ssm_intervals} and figure \ref{boxplot_fig}. No statistically significant differences were found between the two cohorts, except for PPV ($p=0.040$). Nonetheless, PPV mean values only differed by $0.03 \:m/s$ and similar standard deviations were observed (0.12 and 0.08 for clinical and synthetic cohorts, respectively).

\begin{table}[htbp]
\centering
\begin{tabular}{lccc}
\hline
Descriptor & Clinical cohort & Synthetic cohort &   \textit{p} value\\ \hline
PPV [m/s] & $0.59 \pm 0.12$  &  $0.56 \pm 0.08$    & $0.040^\dagger$ \\
FDI [\%]  & $10.36 [5.78, 18.53]$ &  $9.72 [6.30, 16.11]$  & $0.187$ \\
FJA [°]   & $66.77 [27.94, 106.29]$ &  $68.74 [30.09, 101.60]$    &  $0.929$\\
SFD [-]  & $3.84 [0.62, 20.45]$ &  $2.85 [0.65, 15.72]$   & $0.190$ \\
RFI [\%] &  $4.11 [0.00, 32.10]$   & $5.86 [0.00, 27.74]$  & $0.728$ \\
\hline
\end{tabular}
\caption{Comparison of time-averaged flow descriptors between real and synthetic cohorts. Normally distributed variables are expressed as mean $\pm$ standard deviation; non-normally distributed variables are expressed as median [min, max]. $^\dagger$ indicates statistical significance $p \leq 0.05$.}
\label{tab:ssm_intervals}
\end{table}

\subsection{Associations of velocity profile modes with flow morphology descriptors}
For the first four modes, the obtained extreme shape variations are visualized in figure~\ref{profiles_fig}d. Although it is not straightforward to associate each mode of variation with a specific flow feature, it can be hypothesized that the first PCA modes are related to some physically meaningful characteristic of the profile. In practice, the potential physical meaning of each mode was assessed by evaluating equation~\ref{eq_ssm} taking into account only one mode at a time and choosing a set of 10 evenly spaced coefficients $b^{(m)} \in [-3, 3]$. 
The flow morphology descriptors introduced in section~\ref{subsec:descriptors}, were computed for each generated profile as $b^{(m)}$ was gradually increased. For simplicity, correlations with spatial and temporal flow features were analyzed separately. Correlations with spatial features were computed at PS.

\paragraph{Modes correlation with spatial features at peak systole}
Correlation results at PS are reported in table \ref{tab:pca_descr}. 
Mode 1 seemed related to the spatial heterogeneity 
of velocity magnitude and FDI, with larger high velocity regions observed when $b^{(1)} = 3$. This was confirmed by the statistically significant positive correlations of mode 1 with PPV ($r=0.94$, $p<0.0001$) and FDI ($r=0.99$, $p<0.0001$). Overall, with increasing $b^{(1)}$, PPV, FDI, SFD, 
and FJA increased, indicating a tendency of the profile to be less aligned with the plane normal. 
Similar trends were obtained for velocity profiles generated by varying $b^{(2)}$, which appeared related to overall flow rate together with the size of the region with intermediate velocity as it can be visualized by the less pronounced jet for $b^{(2)} = -3$ in figure \ref{profiles_fig}. With increasing $b^{(2)}$, PPV, SFD, 
and FJA significantly increased ($r=0.86$, $p=0.001$; $r>0.99$, $p<0.0001$; $r>0.99$, $p<0.0001$, respectively). No statistically significant correlation was found between mode 2 and FDI ($r=-0.14$, $p=0.702$).
Profiles corresponding to $b^{(3)} = -3$ showed a more spatially concentrated velocity jet than profiles sampled at  $b^{(3)} = 3$. Quantitatively, this was reflected in increasing SFD ($r=0.88$, $p=0.001$) and FJA ($r=0.88$, $p=0.001$) and in decreasing PPV ($r=-0.67$, $p=0.032$) as $b^{(3)}$ increased. No correlation was found between mode 3.
Finally, profiles generated by varying $b^{(4)}$ showed differences in the shape of the high velocity region and in jet orientation. Specifically, with increasing $b^{(4)}$, FJA ($r=-0.91$, $p<0.0001$) and SFD ($r=-0.90$, $p<0.0001$) decreased while FDI increased ($r=0.96$, $p<0.0001$). No correlation was found between mode 4 and PPV 
($r=-0.27$, $p=0.458$).

\begin{table}[htbp]
\resizebox{\textwidth}{!}{\begin{tabular}{lcccccccccc|cc}
\hline
    &       &       &       &       &   $b^{(m)}$    &       &      &       &       &       & $r$     & $p$      \\ 
    & -3.00 & -2.33 & -1.67 & -1.00 & -0.33 & 0.33   & 1.00 & 1.67  & 2.33  & 3.00  &       &        \\ \hline
    
    &       &       &       &       &  Mode 1 &      &      &       &       &       &       &        \\ 
PPV [m/s]& 1.34  & 1.35  & 1.37  & 1.39  & 1.42   & 1.52 & 1.62 & 1.73 & 1.87 & 2.03 & 0.94 & $<0.0001$  \\
FDI [\%] & 7.42  & 8.09  & 8.86  & 9.83  & 11.78   & 14.01 & 15.92 & 17.53 & 18.70 & 19.59 & 0.99 & $<0.0001$      \\
FJA [°] & 9.48  & 11.54  & 13.08  & 14.25  & 12.61   & 13.44 & 14.09 & 14.61 & 15.02 & 15.36 & 0.88 & 0.001 \\
SFD [-]& 0.16  & 0.20  & 0.23  & 0.25  & 0.22   & 0.23 & 0.25 & 0.26 & 0.26 & 0.27 & 0.88 & 0.001        \\ \hline
    
    &       &       &       &       &  Mode 2 &      &      &       &       &       &       &        \\ 
PPV [m/s]& 1.53  & 1.40  & 1.31  & 1.28  & 1.38   & 1.56 & 1.76 & 1.96 & 2.16 & 2.37 & 0.86 & 0.001   \\
FDI [\%]& 8.93  & 9.86  & 11.63  & 14.66  & 13.76   & 12.02 & 11.01 & 10.30 & 9.78 & 9.54 & -0.14 & 0.702   \\
FJA [°] & 7.56  & 8.82  & 10.06  & 11.27  & 12.47   & 13.64 & 14.78 & 15.91 & 17.01 & 18.09 & \textgreater{0.99} & $<0.0001$   \\
SFD [-]& 0.13  & 0.15  & 0.17  & 0.20  & 0.22  & 0.24 & 0.26 & 0.28 & 0.30 & 0.32 & \textgreater{0.99} & $<0.0001$ \\ \hline
    
    &       &       &       &       &  Mode 3 &      &      &       &       &       &       &        \\ 
PPV [m/s]& 2.03  & 1.83  & 1.65  & 1.53  & 1.49   & 1.46 & 1.43 & 1.43 & 1.51 & 1.64 & -0.67 & 0.032 \\
FDI [\%]& 12.40  & 14.16  & 15.33  & 15.25  & 13.38   & 12.40 & 12.75 & 13.45 & 12.94 & 11.29 & -0.52 & 0.122 \\
FJA [°] & 12.86  & 12.79  & 12.78  & 12.82  & 12.95   & 13.19 & 13.56 & 14.09 & 14.82 & 15.78 & 0.88 & 0.001 \\
SFD [-]& 0.22  & 0.22  & 0.22  & 0.22  & 0.23   & 0.23 & 0.24 & 0.25 & 0.26 & 0.28 & 0.88 & 0.001 \\  \hline
    
    &       &       &       &       &  Mode 4 &      &      &       &       &       &       &        \\ 
PPV [m/s]& 1.66  & 1.58  & 1.51  & 1.49  & 1.47   & 1.47 & 1.47 & 1.50 & 1.54 & 1.60 & -0.27 & 0.458 \\
FDI [\%]& 11.60  & 11.50  & 11.43  & 12.21  & 12.31   & 13.14 & 14.12 & 15.09 & 16.08 & 16.52 & 0.96 & $<0.0001$ \\
FJA [°] & 41.68  & 35.70  & 28.65  & 21.83  & 15.74   & 10.66 & 7.13 & 6.26 & 7.90 & 10.40 & -0.91  & $<0.0001$ \\
SFD [-]& 0.89  & 0.71  & 0.54  & 0.40  & 0.28   & 0.18 & 0.12 & 0.10 & 0.13 & 0.18 & -0.90& $<0.0001$ \\ \hline

\end{tabular}}
\caption{PCA modes correlations with velocity profile flow descriptors at peak systole. Results of Pearson correlation analyses are reported. $p<0.05$ indicates statistical significance.}
\label{tab:pca_descr}
\end{table}

\begin{figure}[ht]
\centering
\includegraphics[width=0.8\textwidth]{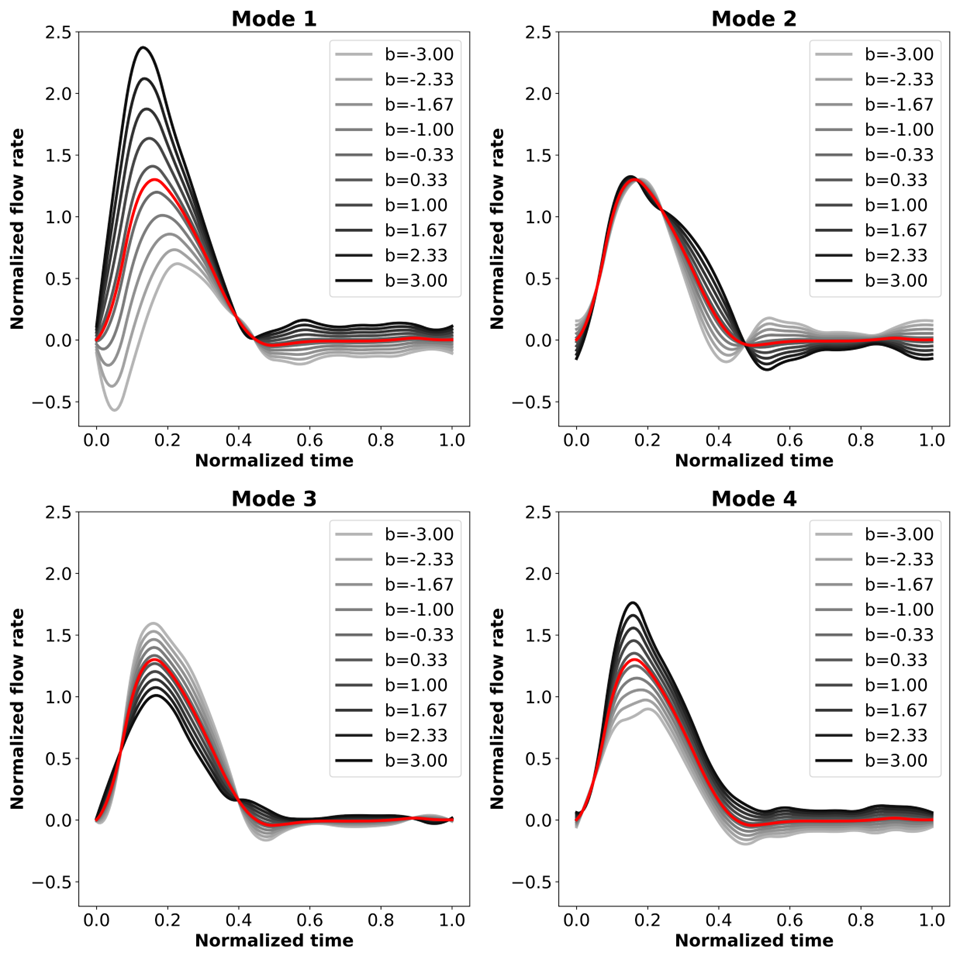}
\caption{Plots of normalized flow rates over time computed for the first four modes by varying their corresponding coefficient $b^{(m)}$. Mean flow rate over time corresponding to $\bar{\mathbf{V}}$ is shown in red for all subplots.} 
\label{flowrates_fig}
\end{figure}

\paragraph{Modes correlation with temporal features}
For each of the first four modes, the flow rates computed on the 10 evenly spaced coefficients $b^{(m)} \in [-3, 3]$ were analyzed (figure \ref{flowrates_fig}). 
Profiles corresponding to mode 1 showed the largest variation in peak flow rate, ranging from 0.62 for $b^{(1)} = -3$ to 2.37 for $b^{(1)} = 3$, and the largest temporal shift ($\Delta t = 0.096$). Furthermore, a negative linear correlation was found between $b^{(1)}$ and RFI ($r=-0.87$, $p=0.001$), with larger portions of retrograde flow corresponding to lower values of $b^{(1)}$.
A similar trend was observed for mode 4, whose associated flow rate peaks increased with increasing $b^{(4)}$, but within a narrower range ([0.90, 1.76]) and with a smaller temporal shift ($\Delta t = 0.04$). 
On the other hand, an opposite trend was found for mode 3, with higher flow rates for decreasing $b^{(3)}$, and with negligible temporal shifts.
Flow rates associated to profiles generated by varying $b^{(2)}$ all showed similar flow rate curves over time.

\clearpage
\section{Discussion}
\label{sec:discussion}

The lack of patient-specific information on \textit{in vivo} flow features has led to the widespread use of simplified inlet velocity profile BCs when setting up CFD models. However, this assumption can significantly impact results, particularly when modeling the ascending aorta \citep{morbiducci2013inflow, pirola2018computational, youssefi2018impact}. The present work addressed the issue of scarcity of patient-specific hemodynamic data by proposing a valid alternative to idealized inlet velocity profiles. In particular, we focused on cases of ATAA.

The first key achievement of the present work was exploiting SSM to generate a dataset of 437 synthetic velocity profiles starting from a clinical cohort of 30 ATAA subjects. The proposed methodology allowed to create aortic inlet velocity fields that presented similar spatiotemporal characteristics as compared to real ATAA patients, and that are suitable to be prescribed as inlet BC for CFD simulations. We found an average velocity profile $\bar{\mathbf{V}}$ that qualitatively resembled a 3D paraboloid-like shape (figure \ref{profiles_fig}c). This finding suggests that imposing an idealized parabolic profile as inflow BC of an ATAA CFD model is the best choice in the absence of patient-specific flow data. Similar findings were reported by \citep{youssefi2018impact}, who also suggested that a parabolic profile is a reasonable choice when patient-specific data are missing and in case of TAV. Nonetheless, our $\bar{\mathbf{V}}$ was characterized by $FJA=13$\textdegree and $SFD=0.23$ at PS, whereas a perfectly symmetric and centered paraboloid would have null $FJA$ and $SFD$ values. In particular, the velocity jet was directed toward the right side of the subject, i.e., the extrados of the ascending aorta. As recognized by \citep{youssefi2018impact} and confirmed by additional studies (\citep{pirola2018computational, armour2021influence}), including in-plane velocity components significantly affects flow dynamics predictions in the ascending aorta. Therefore, for an accurate assessment of ATAA hemodynamics, the mean profile proposed in this study represents a better baseline choice for inlet BC specification in CFD simulations. \\
Our SSM was also sampled to generate new realistic 4D profiles;
the individual variance associated to the first mode was equal to $\approx 13\%$, whereas the first four modes accounted for $\approx 45\%$ of cumulative variance (figure \ref{profiles_fig}a). These findings significantly deviate from results reported by previous studies involving PCA on aortic geometry \citep{thamsen2021synthetic, cosentino2020statistical}. 
Thamsen \textit{et al.} \citep{thamsen2021synthetic} built an SSM of aortic coarctation anatomy and found that the first mode covered 43.7\% of the total variation, while the first three modes were responsible for its 70\%. The authors also built an SSM of inlet vector fields based on 4D flow MRI data, but did not provide full insights into profile variability and only considered PS, limiting their analysis to steady conditions. 
Our results indicate substantially lower individual mode explained variance. Such discrepancy can be attributed mainly to 
the fact that we dealt with 3D velocity vectors that change over time, which represent considerably more complex, higher dimensional features as compared to 3D positions describing static geometries.

The second key achievement of this study was providing insights into the hemodynamic features that are responsible for the most significant variations of ascending aorta spatiotemporal velocity profiles. Although it was not possible to find a one to one correspondence between shape modes and flow descriptors, we were able to demonstrate that the first four modes can be related to a unique combinations of spatial flow morphology descriptors (table \ref{tab:pca_descr} and figure \ref{profiles_fig}). A more straightforward interpretation of PCA modes has been reported in studies dealing with aortic anatomy \citep{casciaro2014identifying, cosentino2020statistical}. In our results, an overlapping of feature contributions to PCA modes is likely due to the complexity and high dimensionality of our 4D velocity vector fields that makes our data significantly more heterogeneous. Nonetheless, it is interesting to note that mode 1 was clearly linked to a temporal feature: flow rate over time (figure \ref{flowrates_fig}). Spatially, this resulted in a positive correlation between $b^{(1)}$ and systolic $PPV$ ($r=0.94$, $p<0.0001$) and $FDI$ ($r=0.99$, $p<0.0001$), which both contribute to increase net flow rate. Therefore, when exploiting the proposed SSM, users that wish to investigate the impact of specific flow features will be able to adjust mode weights to generate profiles with the desired characteristics. 

\section{Conclusions}
In this work we built the first data-driven generative model of time-dependent 3D aortic velocity profiles, suitable to be used in numerical simulations of blood flow. With the aim of expediting the development of future \textit{in silico} analyses, the proposed software system also allows to map any of the generated velocity profiles to the inlet plane of any virtual subject given its coordinate set. The present work thus sets a new standard for the computational bioengineering community, allowing to replace the common practice of prescribing idealized inflow BC in numerical simulations of blood flow with more realistic conditions.

\section*{Data availability and usability}
We provide the generated synthetic cohort of 4D velocity profiles ready to be used for time-dependent CFD simulations with a Github repository also containing the necessary scripts to replicate our study. The code can be found at: \url{https://github.com/saitta-s/flow4D} and all synthetic velocity profiles can be downloaded from \url{https://doi.org/10.5281/zenodo.7251987}.

\section*{Acknowledgements}
This work was supported by the National Institute for Health Research (NIHR) Imperial College Biomedical Research Centre (P69559) and the Imperial College London British Heart Foundation Centre for Research Excellence (RG/19/6/34387, RE/18/4/34215). SP acknowledges the support of the Delft Technology Fellowship. DO'R is supported by the Medical Research Council (MC\textunderscore UP \textunderscore 1605/13).

For the purpose of open access, the authors have applied a creative commons attribution (CC BY) licence to any author accepted manuscript version arising.

\clearpage
\bibliographystyle{elsarticle-num} 
\bibliography{biblio}

\begin{thebibliography}{10}
\expandafter\ifx\csname url\endcsname\relax
  \def\url#1{\texttt{#1}}\fi
\expandafter\ifx\csname urlprefix\endcsname\relax\def\urlprefix{URL }\fi
\expandafter\ifx\csname href\endcsname\relax
  \def\href#1#2{#2} \def\path#1{#1}\fi

\bibitem{elefteriades2002natural}
J.~A. Elefteriades, Natural history of thoracic aortic aneurysms: indications
  for surgery, and surgical versus nonsurgical risks, The Annals of thoracic
  surgery 74~(5) (2002) S1877--S1880.

\bibitem{catalano2021atlas}
C.~Catalano, V.~Agnese, G.~Gentile, G.~M. Raffa, M.~Pilato, S.~Pasta,
  Atlas-based evaluation of hemodynamic in ascending thoracic aortic aneurysms,
  Applied Sciences 12~(1) (2021) 394.

\bibitem{yeung2006aortoiliac}
J.~J. Yeung, H.~J. Kim, T.~A. Abbruzzese, I.~E. Vignon-Clementel, M.~T.
  Draney-Blomme, K.~K. Yeung, I.~Perkash, R.~J. Herfkens, C.~A. Taylor, R.~L.
  Dalman, Aortoiliac hemodynamic and morphologic adaptation to chronic spinal
  cord injury, Journal of vascular surgery 44~(6) (2006) 1254--1265.

\bibitem{chien1998effects}
S.~Chien, S.~Li, J.~Y. Shyy, Effects of mechanical forces on signal
  transduction and gene expression in endothelial cells, Hypertension 31~(1)
  (1998) 162--169.

\bibitem{pirola20194}
S.~Pirola, B.~Guo, C.~Menichini, S.~Saitta, W.~Fu, Z.~Dong, X.~Y. Xu, 4-d flow
  mri-based computational analysis of blood flow in patient-specific aortic
  dissection, IEEE Transactions on Biomedical Engineering 66~(12) (2019)
  3411--3419.

\bibitem{mendez2018comparison}
V.~Mendez, M.~Di~Giuseppe, S.~Pasta, Comparison of hemodynamic and structural
  indices of ascending thoracic aortic aneurysm as predicted by 2-way fsi, cfd
  rigid wall simulation and patient-specific displacement-based fea, Computers
  in biology and medicine 100 (2018) 221--229.

\bibitem{li2005blood}
Z.~Li, C.~Kleinstreuer, Blood flow and structure interactions in a stented
  abdominal aortic aneurysm model, Medical engineering \& physics 27~(5) (2005)
  369--382.

\bibitem{peirlinck2021precision}
M.~Peirlinck, F.~S. Costabal, J.~Yao, J.~Guccione, S.~Tripathy, Y.~Wang,
  D.~Ozturk, P.~Segars, T.~Morrison, S.~Levine, et~al., Precision medicine in
  human heart modeling, Biomechanics and modeling in mechanobiology 20~(3)
  (2021) 803--831.

\bibitem{morbiducci2013inflow}
U.~Morbiducci, R.~Ponzini, D.~Gallo, C.~Bignardi, G.~Rizzo, Inflow boundary
  conditions for image-based computational hemodynamics: impact of idealized
  versus measured velocity profiles in the human aorta, Journal of biomechanics
  46~(1) (2013) 102--109.

\bibitem{pirola2018computational}
S.~Pirola, O.~Jarral, D.~O'Regan, G.~Asimakopoulos, J.~Anderson, J.~Pepper,
  T.~Athanasiou, X.~Xu, Computational study of aortic hemodynamics for patients
  with an abnormal aortic valve: The importance of secondary flow at the
  ascending aorta inlet, APL bioengineering 2~(2) (2018) 026101.

\bibitem{youssefi2018impact}
P.~Youssefi, A.~Gomez, C.~Arthurs, R.~Sharma, M.~Jahangiri,
  C.~Alberto~Figueroa, Impact of patient-specific inflow velocity profile on
  hemodynamics of the thoracic aorta, Journal of biomechanical engineering
  140~(1) (2018).

\bibitem{armour2021influence}
C.~H. Armour, B.~Guo, S.~Pirola, S.~Saitta, Y.~Liu, Z.~Dong, X.~Y. Xu, The
  influence of inlet velocity profile on predicted flow in type b aortic
  dissection, Biomechanics and modeling in mechanobiology 20~(2) (2021)
  481--490.

\bibitem{romero2021clinically}
P.~Romero, M.~Lozano, F.~Mart{\'\i}nez-Gil, D.~Serra, R.~Sebasti{\'a}n,
  P.~Lamata, I.~Garc{\'\i}a-Fern{\'a}ndez, Clinically-driven virtual patient
  cohorts generation: An application to aorta, Frontiers in Physiology (2021)
  1375.

\bibitem{young2009computational}
A.~A. Young, A.~F. Frangi, Computational cardiac atlases: from patient to
  population and back, Experimental physiology 94~(5) (2009) 578--596.

\bibitem{casciaro2014identifying}
M.~E. Casciaro, D.~Craiem, G.~Chironi, S.~Graf, L.~Macron, E.~Mousseaux,
  A.~Simon, R.~L. Armentano, Identifying the principal modes of variation in
  human thoracic aorta morphology, Journal of Thoracic Imaging 29~(4) (2014)
  224--232.

\bibitem{jollife2016principal}
I.~T. Jollife, J.~Cadima, Principal component analysis: A review and recent
  developments, Philosophical Transactions of the Royal Society A:
  Mathematical, Physical and Engineering Sciences 374~(2065) (2016) 20150202.

\bibitem{liang2017machine}
L.~Liang, M.~Liu, C.~Martin, J.~A. Elefteriades, W.~Sun, A machine learning
  approach to investigate the relationship between shape features and
  numerically predicted risk of ascending aortic aneurysm, Biomechanics and
  modeling in mechanobiology 16~(5) (2017) 1519--1533.

\bibitem{thamsen2021synthetic}
B.~Thamsen, P.~Yevtushenko, L.~Gundelwein, A.~A. Setio, H.~Lamecker, M.~Kelm,
  M.~Schafstedde, T.~Heimann, T.~Kuehne, L.~Goubergrits, Synthetic database of
  aortic morphometry and hemodynamics: overcoming medical imaging data
  availability, IEEE Transactions on Medical Imaging 40~(5) (2021) 1438--1449.

\bibitem{cosentino2020statistical}
F.~Cosentino, G.~M. Raffa, G.~Gentile, V.~Agnese, D.~Bellavia, M.~Pilato,
  S.~Pasta, Statistical shape analysis of ascending thoracic aortic aneurysm:
  correlation between shape and biomechanical descriptors, Journal of
  Personalized Medicine 10~(2) (2020) 28.

\bibitem{nannini2021aortic}
G.~Nannini, A.~Caimi, M.~C. Palumbo, S.~Saitta, L.~N. Girardi, M.~Gaudino,
  M.~J. Roman, J.~W. Weinsaft, A.~Redaelli, Aortic hemodynamics assessment
  prior and after valve sparing reconstruction: A patient-specific 4d
  flow-based fsi model, Computers in Biology and Medicine 135 (2021) 104581.

\bibitem{salmasi2021high}
M.~Y. Salmasi, S.~Pirola, S.~Sasidharan, S.~M. Fisichella, A.~Redaelli, O.~A.
  Jarral, D.~P. O’Regan, A.~Y. Oo, J.~E. Moore~Jr, X.~Y. Xu, et~al., High
  wall shear stress can predict wall degradation in ascending aortic aneurysms:
  an integrated biomechanics study, Frontiers in Bioengineering and
  Biotechnology (2021) 935.

\bibitem{pirola2017choice}
S.~Pirola, Z.~Cheng, O.~Jarral, D.~O'Regan, J.~Pepper, T.~Athanasiou, X.~Xu, On
  the choice of outlet boundary conditions for patient-specific analysis of
  aortic flow using computational fluid dynamics, Journal of biomechanics 60
  (2017) 15--21.

\bibitem{yushkevich2017itk}
P.~A. Yushkevich, G.~Gerig, Itk-snap: an intractive medical image segmentation
  tool to meet the need for expert-guided segmentation of complex medical
  images, IEEE pulse 8~(4) (2017) 54--57.

\bibitem{saitta2022deep}
S.~Saitta, F.~Sturla, A.~Caimi, A.~Riva, M.~C. Palumbo, G.~Nano, E.~Votta,
  A.~D. Corte, M.~Glauber, D.~Chiappino, et~al., A deep learning-based and
  fully automated pipeline for thoracic aorta geometric analysis and planning
  for endovascular repair from computed tomography, Journal of Digital Imaging
  35~(2) (2022) 226--239.

\bibitem{saitta2021qualitative}
S.~Saitta, B.~Guo, S.~Pirola, C.~Menichini, D.~Guo, Y.~Shan, Z.~Dong, X.~Y. Xu,
  W.~Fu, Qualitative and quantitative assessments of blood flow on tears in
  type b aortic dissection with different morphologies, Frontiers in
  Bioengineering and Biotechnology 9 (2021).

\end{thebibliography}





\end{document}